\documentclass[aps,prd,nofootinbib,twocolumn,floatfix]{revtex4-1}

\usepackage{graphicx}
\usepackage{dcolumn}
\usepackage{subfigure}
\usepackage{times,mathptm}
\usepackage{float}
\usepackage{color}
\usepackage{amsmath,amsfonts}
\usepackage{mathptmx}
\usepackage{mathrsfs}
\usepackage{bbm}
\usepackage{bm}
\usepackage{xfrac}
\DeclareMathOperator{\Det}{Det}

\newcommand{\beq}{\begin{eqnarray}}
\newcommand{\eeq}{\end{eqnarray}}

\newcommand{\by}{\mathbf{y}}
\newcommand{\bc}{\mathbf{c}}
\newcommand{\bA}{\mathbf{A}}
\newcommand{\bB}{\mathbf{B}}
\newcommand{\bx}{\mathbf{x}}
\newcommand{\be}{\mathbf{e}}
\newcommand{\bk}{\mathbf{k}}
\newcommand{\bq}{\mathbf{q}}

\usepackage{ulem}

\def\lsim{\mathrel{\rlap{\lower4pt\hbox{\hskip1pt$\sim$}}
    \raise1pt\hbox{$<$}}}
\def\gsim{\mathrel{\rlap{\lower4pt\hbox{\hskip1pt$\sim$}}
    \raise1pt\hbox{$>$}}}
\begin{document}
\bibliographystyle{h-physrev5}

\title{Gluon chain formation in presence of static charges}

\author{
A. Ostrander$^{1,2}$,  E.~Santopinto$^3$, A. P.~Szczepaniak$^2$, A.Vassallo$^3$ }
\affiliation{
$^1$ Department of Physics, Astronomy, and Geology, Berry College, Mount Berry, GA 30149 USA \\
$^2$ Physics Department and Center for Exploration of Energy and Matter, Indiana University, Bloomington, IN 47403 USA\\ 
$^3$ INFN, Sezione di Genova, via Dodecaneso 33, 16146 Genova, Italy \\ } 
\date{\today}
\begin{abstract}
We consider the origins of the gluon chain model. The model  serves as a realization of the dynamics of the 
chromoelectric flux between static quark-antiquark sources. The derivation is based on the large-$N_C$ limit
  of the Coulomb gauge Hamiltonian in the presence of a background field introduced to model magnetic confinement. 
  \end{abstract}

%
%

%
\maketitle

\section{\label{sec:intro}Introduction}

The gluon chain model of Greensite and Thorn~\cite{Greensite:2001nx,Greensite:2002bh,Brower:2006hf,Greensite:2009mi} identifies the chromoelectric  flux tube that exists between static quark charges with a string of quasi particles, constituent gluons. Through lattice simulations and phenomenological analyses it is well established that the instantaneous, Coulomb  potential between static charges is 
confining~\cite{Greensite:2003xf,Nakagawa:2006fk,Voigt:2008rr,Burgio:2012bk}.  Even though it  does not correspond to a physical observable, the static potential does provide physical insight into the possible origins of the confinement mechanism as illustrated by the Gribov-Zwanzinger model~\cite{Gribov:1977wm,Zwanziger:1995cv} and other,  {\it e.g.} variational 
 models~\cite{Szczepaniak:2001rg,Szczepaniak:2003ve,Feuchter:2004gb,Feuchter:2004mk,Epple:2007ut}.   Lattice simulations indicate that the corresponding string tension is larger (by a factor of 2 to 3) as compared to the string tension extracted from time-dependent  large Wilson loops. This is consistent with expectations of  variational analysis.  At fixed quark-antiquark separation the Coulomb potential corresponds to the energy of a quark-antiquark pair in a vacuum state that is unmodified by the presence of the  pair while the energy extracted from the Wilson loop corresponds to the energy of the exact QCD eigenstate in which the quark-antiquark ($Q\bar Q$)  pair polarizes the vacuum~\cite{Zwanziger:2002sh}. The gluon chain model is a particular realization of the latter, {\it i.e.}, the exact pair state.  Confinement originates from the condensation  of chromomagnetic  
 charges~\cite{Del Debbio:1996mh,Langfeld:1997jx,Engelhardt:1999fd,Greensite:2003bk}. Formation of the gluon chain should therefore  also provide  insights into the interplay between constituent gluons and magnetic domains in the vacuum.

 In the Hamiltonian formulation the true $Q\bar Q$  state is generated by the evolution operator $\lim_{\beta \to \infty} \exp(-\beta H)$ from the unperturbed 
  vacuum. This is because in a physical gauge the Hamiltonian $H$ contains all gluon interactions 
   which  also couple  to the  classical, external   quark-antiquark color source. 
 In this paper we investigate if/how the gluon chain emerges from the  evolution operator. 
  We follow a canonical 
  formulation of QCD in the Coulomb gauge since it contains 
   only physical degrees of freedom, and these can be directly related to quasi particles. 
  The gluon field is decomposed into normal modes representing  particle  excitations, and a   physical state is represented as a superposition of multi-gluon states.  Furthermore the normal mode expansion can be performed with respect to
   a non-vanishing  classical background. Such a background is introduced to 
     (phenomenologically) parametrize topologically disconnected sectors of the vacuum. In terms of  the path integral representation these sectors correspond to large field  configurations, {\it i.e.}, field domains that cannot be smoothly connected  to the null field configuration~\cite{Jackiw:1977ng}. 
    
    The paper is organized as follows. In the next section we review the structure of the Hamiltonian,  introduce the particle basis, and discuss the role of the   individual interaction  terms 
     in the formation of the chain.  In Sec.~\ref{com}  we propose a simplified computational  scheme for  studying the formation of  the chain state and discuss numerical results. A summary and outlook are given in Section~\ref{last}.

 \section{QCD Hamiltonian and gluons} 
 
In the Coulomb gauge~\cite{Christ:1980ku} the gluon field  is described by the vector potential,  $\bA^a(\bx)$  that, 
  for each color component $a =1 \cdots N_C^2 - 1$, 
 satisfies the transversality  condition, $\bm{\nabla}\cdot \bA^a=0$. 
 In the Schr\"odinger representation the conjugate momenta, which are proportional to the electric field,  are given by 
  $\bm\Pi^a(\bx) = -i \delta/\delta \bA^a(\bx)$. 
 The  temporal component of the gluon field is eliminated 
  using Gauss's law. This  leads to an instantaneous interaction between color charges. The total color charge density 
  has two components,  $\rho(\bx,a) = \rho_g(\bx,a) + \rho_q(\bx,a)$,  corresponding 
   to gluons and quarks, respectively.  In the following we ignore dynamical quarks, and the only quark charge we consider is that of a  static quark-antiquark pair placed along the $z$-axis a distance $R$ apart. The corresponding density is therefore given by
  \begin{equation} 
\rho_q(\bx,a) =  Q^\dag_i(\bx) T^a_{ij} Q_j(\bx) - \bar Q^\dag_i(\bx)  T^a_{ji} \bar Q_j(\bx).  \label{rhoq} 
\end{equation}
Here $Q^\dag_i(\bx) (Q_i(\bx))$ represents an operator that creates  (annihilates)   a quark at 
 $\bx$ in a state with color $i=1\cdots N_C$, and $T^a$ are the $SU(N_C)$ color matrices in the fundamental representation.  We suppress the (irrelevant) spin indices. 
 Similarly  $\bar Q^\dag_i(\bx)$ and $\bar Q_i(\bx)$ are the creation and annihilation operators for  antiquarks. A state with a static $Q\bar Q$ pair is created by the operator 
$ Q^{\dag}_i( \hat z R/2 ) \bar  Q^\dag_j(- \hat z R/2 )$.  The gluon charge density is given by 
 \begin{equation} 
 \rho_g(\bx,a) =  f_{abc} \bA^b(\bx) \cdot \bm\Pi^c(\bx) ,
 \end{equation} 
 and the Hamiltonian takes the form 
 \begin{equation} 
 H = H_K + H_B + H_C
 \end{equation} 
 where the kinetic plus magnetic terms are given by 
 \begin{equation}
 H_K + H_B  = \frac{1}{2} \int d\bx ({\cal J}^{-1}[\bA] \bm\Pi {\cal J}^{-1}[\bA] \bm\Pi + \bB^2)  ,
 \end{equation}
and 
 \begin{equation} 
H_C = \frac{1}{2} \int d\bx d\by {\cal J}^{-1}[\bA] \rho(\bx,a) {\cal J}[\bA] K^{ab}(\bx,\by,[\bA]) \rho(\by,a)  \label{c} 
 \end{equation} 
 represents the instantaneous Coulomb interaction between color charges. 
Here  ${\cal J}[\bA] = \Det(-{\bf D} \cdot \bm\nabla)$ is the Faddeev-Popov determinant, 
${\bf D} = {\bf D}_{ab} = \bm\nabla \delta_{ab} + g f_{acb} \bA^c $ is  the covariant derivative, and 
$\bB= \bB^a = \nabla \times \bA^a  + g f_{abc} \bA^b \times \bA^c/2$ is the magnetic field. The non-abelian Coulomb kernel is formally given by 
\begin{equation} 
K(\bx,\by,[\bA]) = ({\bf D} \cdot \bm\nabla)^{-1} (-g^2 \bm\nabla^2)  ({\bf D} \cdot \bm\nabla)^{-1}. \label{K}
\end{equation}
The above describes the Hamiltonian in the Schr\"odinger representation. The particle basis 
 representation is obtained  via a canonical transformation from $\bA,\bm\Pi$ to a set of operators $\alpha^\dag(\bk,\lambda,a), \alpha(\bk,\lambda,a)$  representing creation and  
 annihilation of gluons with three-momentum $\bk$ ($k = |\bk|$, $[d\bk] = d\bk/(2\pi)^3$), helicity $\lambda$, and color $a$ 
 \begin{eqnarray} 
 \bA^a(\bx) & = & \int [d\bk]  \frac{1}{\sqrt{2\omega(k)}} [
 \sum_{\lambda=\pm} \be(\bk,\lambda) \alpha(\bk,\lambda,a) e^{i\bk\cdot\bx} + h.c]  \nonumber \\
 \bm\Pi^a(\bx) & = & \frac{1}{i} \int [d\bk] \sqrt{\frac{\omega(k)}{2}} [\sum_{\lambda=\pm}\be(\bk,\lambda) \alpha(\bk,\lambda,a) e^{i\bk\cdot\bx} - h.c].  \nonumber \\ 
 \label{part} 
\end{eqnarray} 
Particle operators satisfy ladder algebra and generate a Fock space labeled by the number of gluons, $n_i$, occupying a state of a given momentum, helicity and color,  $i=(\bk,\lambda,a)$ 
 \begin{equation} 
 |n_1, n_2, \cdots n_i \cdots \rangle = (\alpha_1^\dag)^{n_1} (\alpha_2^\dag)^{n_2} \cdots (\alpha_i^\dag)^{n_i} \cdots  |0\rangle. 
 \end{equation}
  The state with no  gluons,  $|0\rangle \equiv  |0,0,\cdots \rangle$ is annihilated by all annihilation operators $\alpha_i$.

 \subsection{The vacuum state} 
 
 In the absence of quark sources, after normal-ordering the gluon operators, the  Hamiltonian
    \begin{equation} 
  H = \langle 0| H |0 \rangle + :H: \label{no}
  \end{equation} 
 contains an infinite number of  terms that connect states with any numbers of  gluons~\cite{Szczepaniak:2001rg}. 
           The ground  state, $|\Omega\rangle$,   can therefore be formally  written as 
  \begin{equation} 
  |\Omega\rangle =  \left[ \sum_{n_1} \sum_{n_2} \cdots\right]    \Psi_{n_1,n_2,\cdots } |n_1,n_2,\cdots \rangle.  \label{vvv} 
  \end{equation}  
  The non-uniqueness associated with the definition of a gluon state,  and  thus the Hamiltonian 
  in Eq.~(\ref{no}),  arises from the arbitrariness in the choice of the function $\omega(k)$ in Eq.~(\ref{part}).  For example the choice $\omega(k) = k$ corresponds to a basis of non-interacting particles 
    which diagonalizes the  free  Hamiltonian ({\it i.e.}, for $g=0$). Other proposals, based on the  variational principle, have been analyzed  in ~\cite{Szczepaniak:2001rg,Szczepaniak:2003ve,Feuchter:2004gb,Feuchter:2004mk,Epple:2007ut}. 
     These studies considered an optimal choice for the basis of states obtained with $\omega(k)$ that approaches the free particle limit for large $k$ and is large and possibly divergent in the infrared (IR), {\it i.e.},  for $k\to 0$.
This is because an IR enhanced $\omega(k)$ suppresses 
   contributions to  vacuum expectation values ({\it vev}) 
    from fields near the Gribov horizon~\cite{Cucchieri:1996ja} and 
   removes the Landau pole from the  Coulomb kernel  ({\it cf}. Eq.~(\ref{K})).  
  With such an optimal choice
 the vacuum in Eq.~(\ref{vvv}) is approximated 
 by the state with a vanishing number of gluons\footnote{More accurate approximations, which take into account residual correlations among the "optimal gluons," can  be constructed using the standard many-body techniques  of cluster expansion~\cite{Szczepaniak:2002ir,Campagnari:2010wc}.},  {\it i.e.}, $|\Omega \rangle =  |0 \rangle$,    and the ground state energy is therefore given by  the first term in Eq.~(\ref{no}).

\subsection{ The variational $Q\bar Q$ state} 
\label{vqq} 
We next  consider a state containing the $Q\bar Q$ pair. A variational state, $|R\rangle$,  
 which does not  take into account the back reaction of quarks on the vacuum  
  can be defined as (in the volume ${\cal V}$) 
  \begin{equation} 
|R\rangle = \frac{1}{{\cal V}\sqrt{N_C}} Q^{\dag}_i(\frac{R}{2} \hat z) \bar Q^{\dag}_i(-\frac{R}{2} \hat z) |0\rangle,  \label{qqv} 
\end{equation} 
 and it is normalized, $\langle R|R\rangle=1$. 
   Even if $|0\rangle$ was the exact ground state, this state would only be an approximation to the exact QCD eigenstate containing the $Q\bar Q$ pair. 
This is because with   $\rho_q \ne 0$ the term in $H_C$  proportional to $\rho_q \times \rho_g$ 
  does not conserve the gluon number. The expectation value of the Hamiltonian in the variational $Q\bar Q$ state defines the Coulomb potential, $V_c(R)$,  which is proportional to 
  the expectation value of the Coulomb kernel in the variational vacuum,

  \begin{equation} 
V_c(R) \delta_{ab}  =-  \langle 0| K^{ab}(R)|0\rangle. \label{vcc}
\end{equation} 
Here $K^{ab}(R)$ is given by Eq.~(\ref{K}) evaluated at the positions of the quark and the 
 antiquark.  The vacuum expectation value may be  computed by expanding the  covariant derivatives   in  powers of $\bA$ ({\it cf}. Eq.~(\ref{c})) and noticing that in the variational vacuum,  
    \begin{equation} 
 \langle 0| \bA^a(\bx) \bA^b(0)|0 \rangle = \delta_{ab} \int \frac{d\bk}{(2\pi)^3} 
    \frac{\delta_T(\bk)}{2\omega(k)}   e^{i\bk\cdot \bx} \label{aa}
 \end{equation} 
 where  $\delta_T(\bk) =  \sum_{\lambda = \pm}  e_i (\bk,\lambda) e^*_j(\bk,\lambda) =
 \delta_{ij} - k_ik_j/k^2$.   
   The  behavior of $V_c(R)$ at large-$R$  is correlated  with the IR behavior of $\omega(k)$. While early variational studies indicated that with a proper choice of $\omega(k)$ it would be possible to obtain a confining potential,
     more detailed analyses showed that all solutions are  massive, {\it i.e.}, when transformed to momentum space $V_c(k)$  is always finite in the limit $k\to0$, {\it a.k.a.} non-confining~\cite{Epple:2007ut}.  
     We now believe this is consistent with lattice results. 
 As shown in ~\cite{Greensite:2003xf} the large-$R$ strength of  the Coulomb potential originates 
    from magnetic charges in the vacuum. These are absent 
   in the variational model calculation of Eq.~(\ref{vcc}) that is driven by fields in the  neighborhood of the ${\bf A}=0$ configuration. This is because  magnetic charges are topologically disconnected from the first Gribov 
   region where the expansion applies.  Thus it is likely that the string tension, $\sigma_c$,  of the variational model $V_c(r) \sim \sigma_c r$ should at most only be a fraction of the Coulomb string tension and, more likely, $V_c$ of the variational model ought not to be confining. In the following we further explore these scenarios.

\begin{figure}[t!]
\centerline{\scalebox{0.30}{\includegraphics{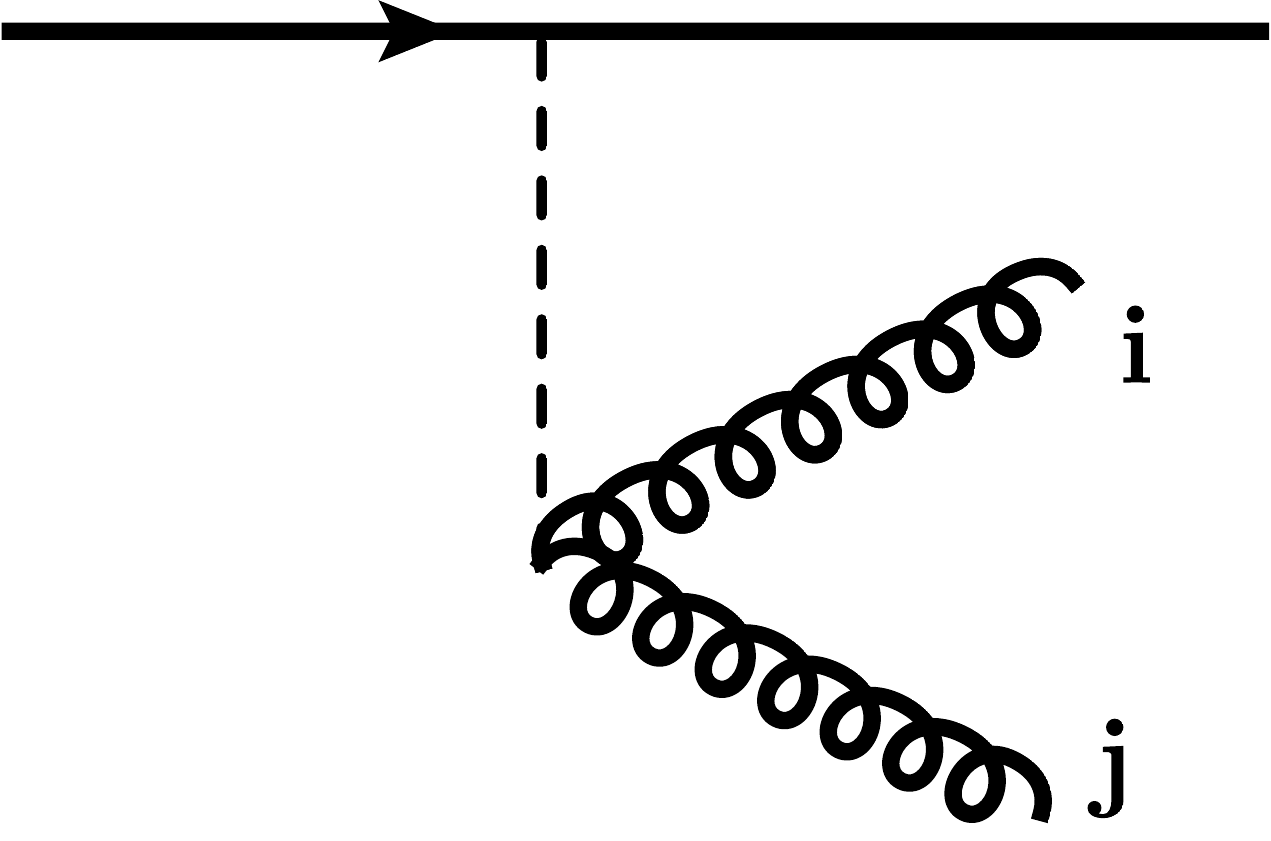}}}
\caption{ Interaction between quark charge (upper line) and the off-diagonal gluon charge $\rho^2_{ij}$ which 
 creates two gluons.  The dashed line represents the Coulomb potential given by the {\it v.e.v} of the fully dressed Coulomb kernel.} 
 \label{gg}
\end{figure}

      It is straightforward to show that 
     the expectation value of the Hamiltonian in the variational $Q\bar Q$ state is given in terms of $V_c$ by 

\begin{equation} 
\langle R|H|R\rangle - \langle 0|H|0\rangle = C_F V_c(R) -  C_F V_c(0)  \label{neq0}
\end{equation} 
where the last term arises  from self-energies of the two static quarks  ($C_F = (N_C^2-1)/2N_C$ is the $SU(N_C)$ color Casimir in the fundamental representation).
As already mentioned above the Coulomb term, $H_C$, involves coupling 
 between quark and gluon charges. It seems reasonable to expect that this interaction might be responsible for generating the  gluon chain.   In the particle basis the gluon charge density is given by 
  \begin{equation} 
   \rho_g(\bx,a)   = \sum_i  \rho^1_i(\bx,a) \alpha^\dag_i \alpha_i + \sum_{ij}  (
   \rho^2_{ij}(\bx,a)   \alpha^\dag_i \alpha^\dag_j 
   + h.c.).  \label{rhog} 
   \end{equation} 
The first term is diagonal in the particle basis and 
 because  $|0\rangle$ contains no gluons
 it  vanishes when applied to the $Q\bar Q$ 
  state defined by  Eq.~(\ref{qqv}).  The second term, however, changes the number of gluons by two 
   and thus could be generating the chain. We will return to this possibility below. 
   There are other, more complicated interactions involving
 the quark charge and gluon operators that  change 
  the number of gluons. They originate  from the $\bA$-dependence of the 
 Coulomb kernel. In the particle basis, the Coulomb kernel can be written as 
\begin{equation} 
K^{ab}(R) = - V_c(R) + :K^{ab}(R): 
\end{equation}
where the normal-ordered part is given by 
\begin{equation} 
:K(R): =   \sum_{\{n\},\{m\} }  K_{n_1,n_2,\cdots;m_1,m_2\cdots}
 (\alpha^{\dag}_1)^{n_1} 
(\alpha^{\dag}_2)^{n_2}  \cdots 
 \alpha_1^{m_1} 
\alpha_2^{m_2}  \cdots . 
\end{equation} 
 Here $K_{\{n\}, \{m\}}$  are the matrix  elements of  the full kernel evaluated between states  containing  $\{n\}$ and $\{m\}$   gluons, respectively. Thus, when multiplied by $\rho_q$ the normal-ordered Coulomb    kernel mixes the variational  $Q\bar Q$ state with states containing arbitrary  numbers   of gluons. 
 As shown in ~\cite{Szczepaniak:2005xi}, however,  in the large-$R$  limit  
the matrix elements $K_{\{n\}, \{m\}}$ for $\{m\},\{n\} \ne 0$ are expected to be smaller than those  for $\{m\}=\{n\}
={0}$. 
  Therefore we expect that at large-$R$  the dominant interaction 
  between quark sources and dynamical gluons  originates from the off-diagonal gluon charge density 
  ({\it c.f.}  Eq.~(\ref{rhog}))  coupled     to the quark charge via $V_c$, and is given by 
  \begin{equation} 
- \int d\bx d\by \rho^a_q(\bx) V_c(|\bx - \by|) \sum_{ij}(\rho^2_{ij}(\by,a) \alpha^\dag_i \alpha^\dag_j  + h.c),  \label{na} 
\end{equation} 
and  shown in Fig.~(\ref{gg}). 
In Eq.~(\ref{na}) the gluon charge density creates (annihilates) two constituent gluons in a color   antisymmetric state. Thus the  combined spin and spacial  wave function of the gluon pair also has to be antisymmetric. However, since $\rho_g$ is a scalar under rotations, the  matrix element, $\rho^2_{ij}$  is symmetric in spin and relative momentum. 
   Thus the above  candidate operator for the gluon chain actually vanishes identically.

The variational basis based on the mode expansion in Eq.~(\ref{part}) seems  incompatible with  the gluon chain picture.  There is further evidence that a model in which the   vacuum is described solely in terms of fluctuations around the $\bA=0$ configuration, as implied by Eq.~(\ref{part}), is inadequate. If  $V_c$ is confining then the expectation value of $H$ in a single gluon state is infinite~\cite{Szczepaniak:1995cw} at all  temperatures, and the model fails to predict the deconfinement  phase transition~\cite{Reinhardt:2011hq}. It is well established  that confinement is related to the presence of magnetic domains in the vacuum, and these are absent in the variational vacuum state. One would expect that the magnetic term $\bB^2$ should play   an important role in confinement  since even the classical Yang-Mills  field equations  have  monopole solutions~\cite{Wu:1967vp}.

\begin{figure}[t!]
\centerline{\scalebox{0.30}{\includegraphics{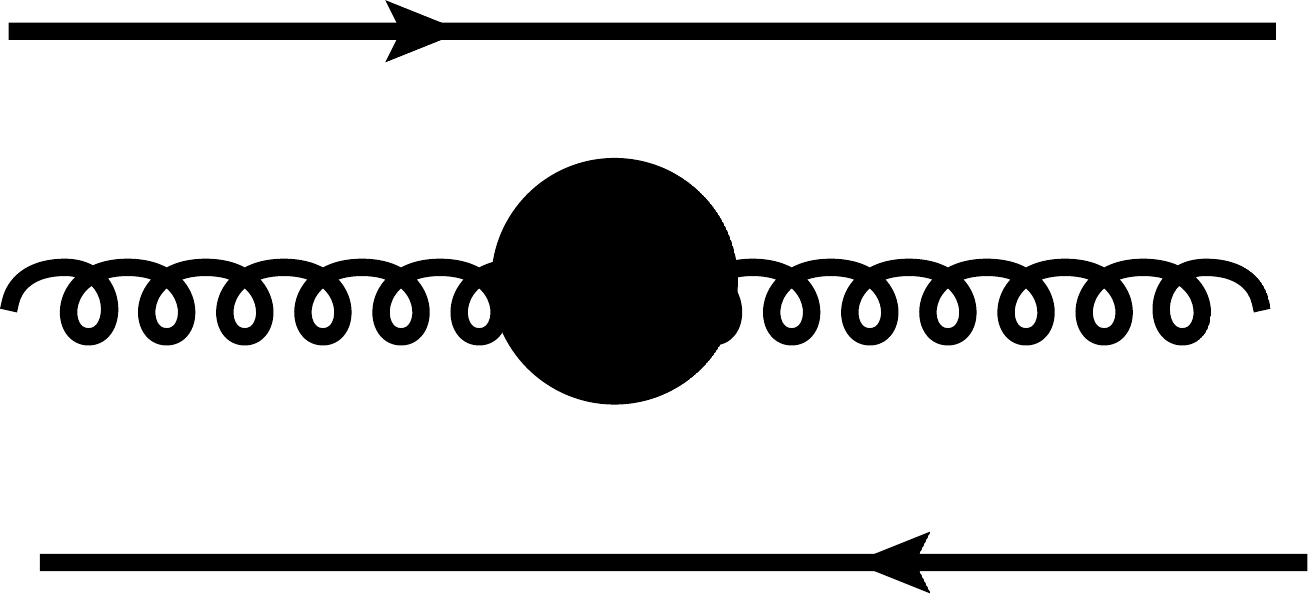}}}
\caption{Illustration of the matrix element of the one-body operator  Hamiltonian, $E_i \alpha^i \alpha_i$, in the gluon chain state with $N=1$. } 
\label{Eq} 
\end{figure}

\begin{figure}[t!]
\centerline{\scalebox{0.30}{\includegraphics{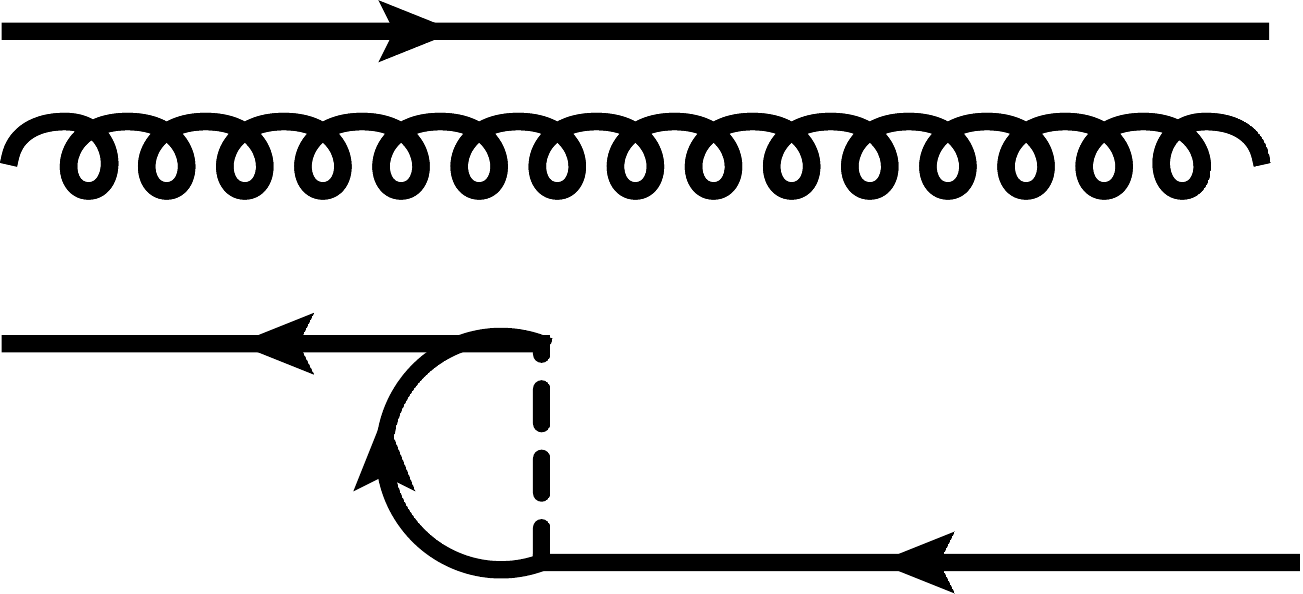}}}
\caption{Part of the leading contribution at large-$N_C$ from $H_{qq}$ which corresponds to quark or antiquark self energy, shown here for the $N=1$ gluon chain state. }
\label{hqq-self} 
\end{figure}

\begin{figure}[t!]
\centerline{\scalebox{0.30}{\includegraphics{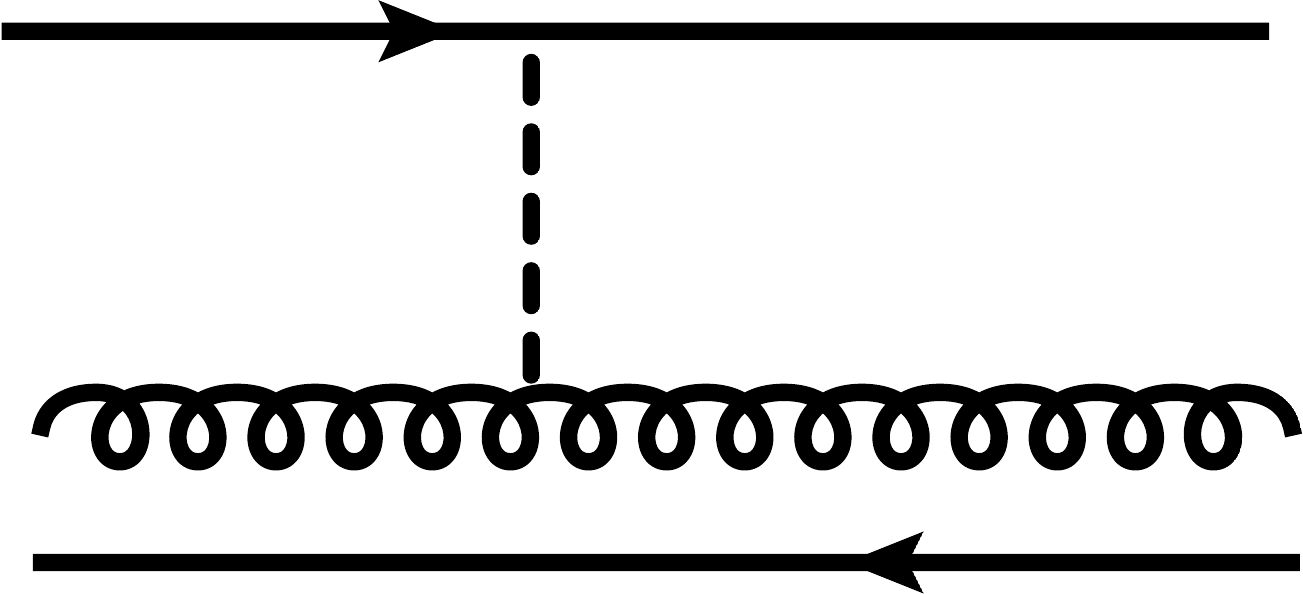}}}
\centerline{\scalebox{0.30}{\includegraphics{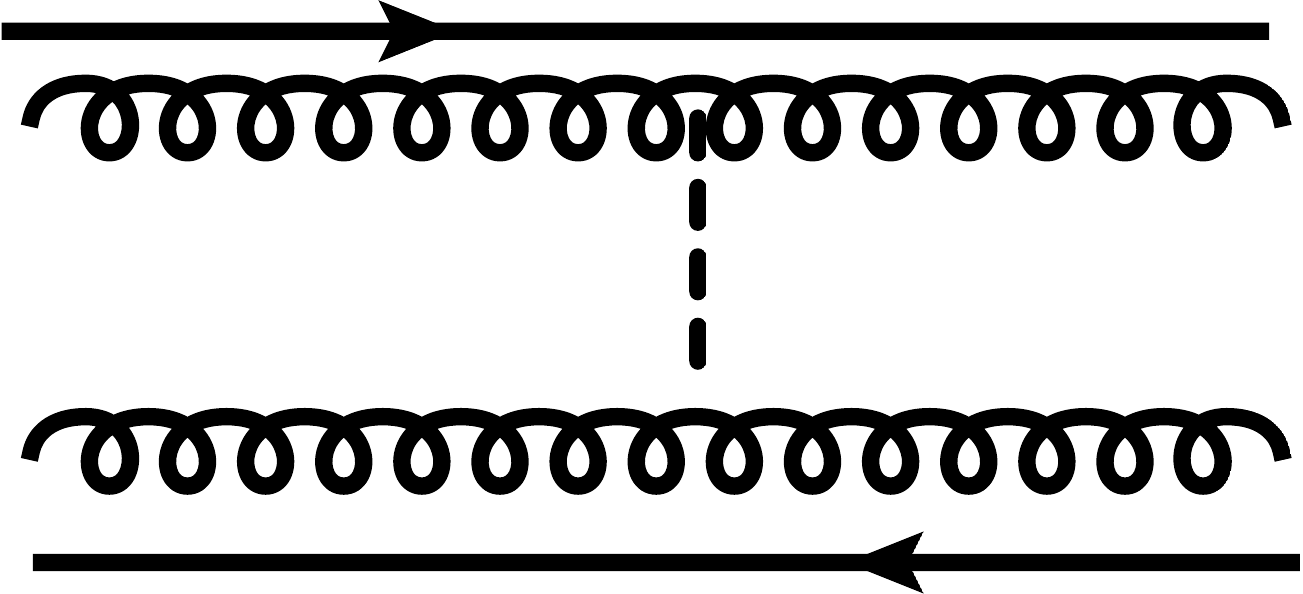}}}
\caption{Quark-gluon, $H^D_{qg}$, and gluon-gluon interaction, $H^D_{gg}$,  matrix elements, for $N=1$ 
and $N=2$ chain states, respectively}
\label{hqq-qg} 
\end{figure}

\begin{figure}[t!]
\centerline{\scalebox{0.30}{\includegraphics{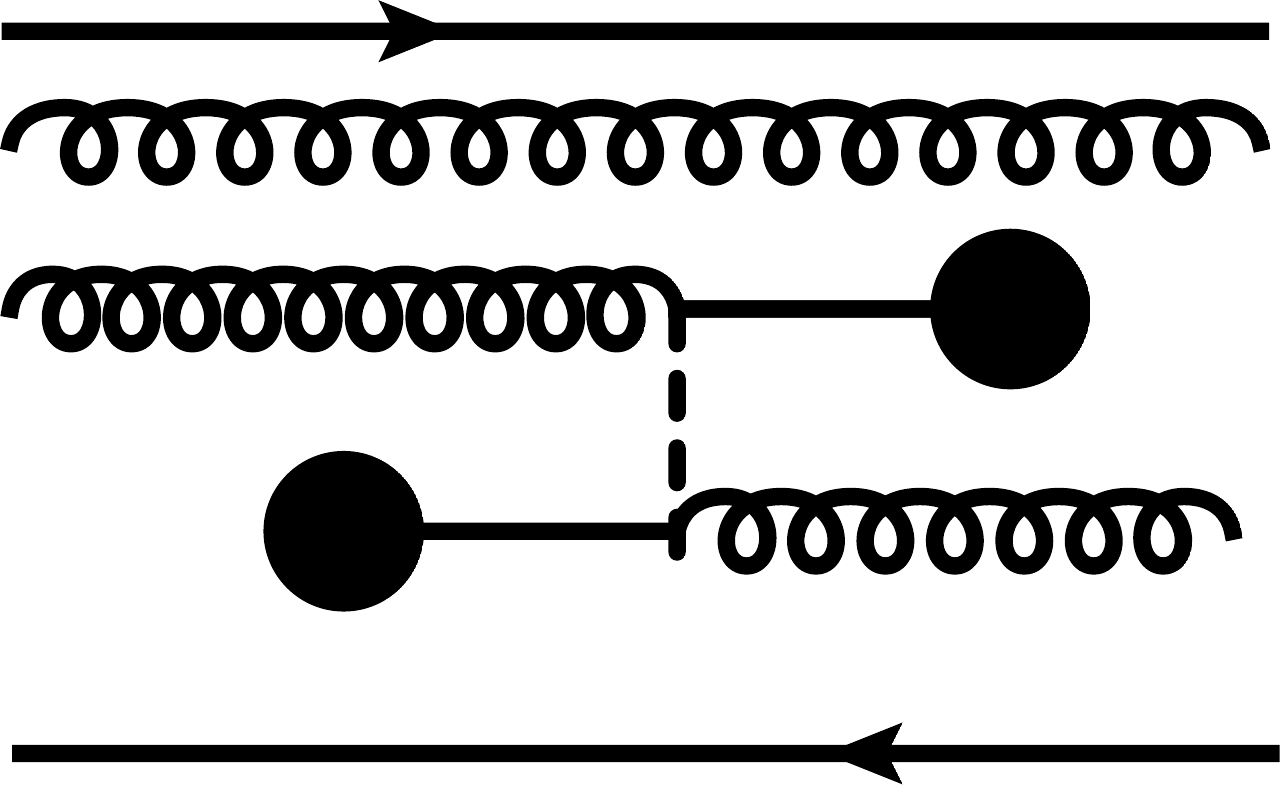}}}
\centerline{\scalebox{0.30}{\includegraphics{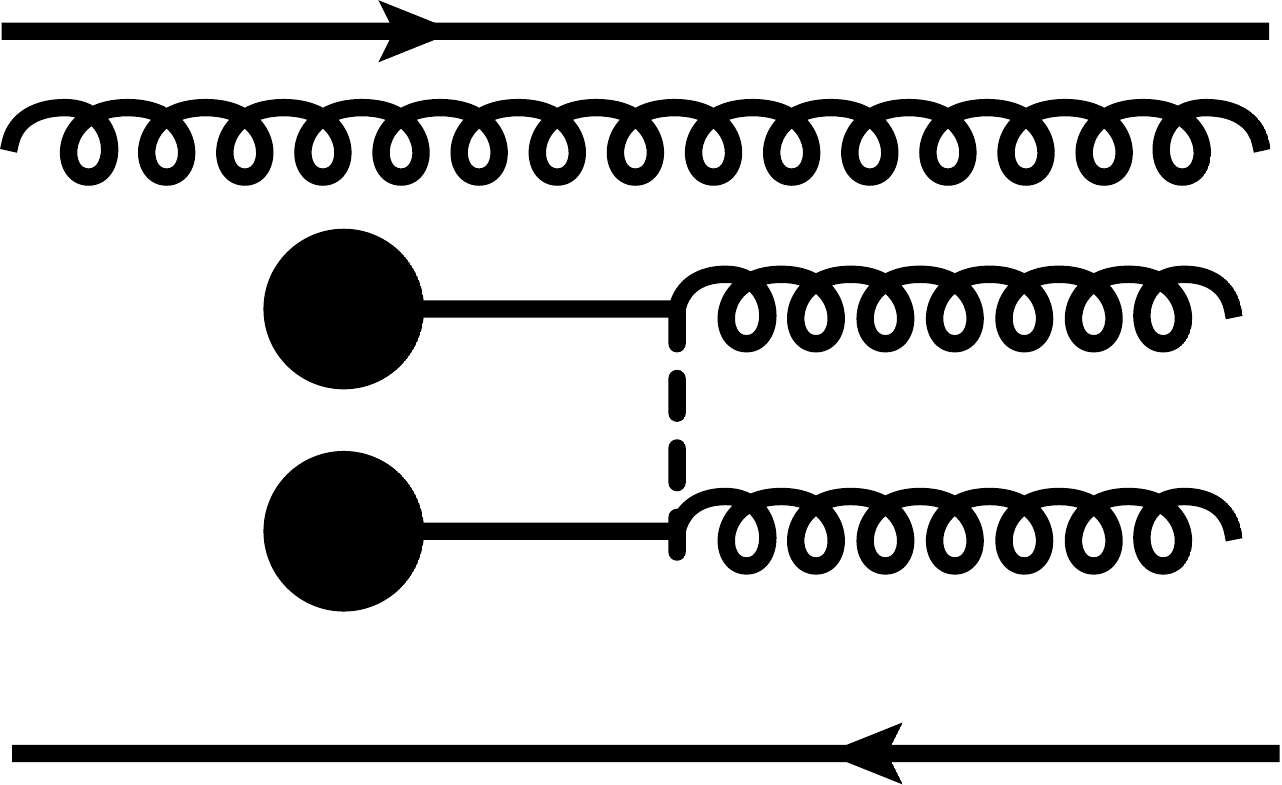}}}
\caption{Diagonal $H^D_{gb}$ and off-diagonal $H^M_{gb}$ interactions between quasi-gluons and the background 
field  (shown as blobs), for $N=3$ gluon chain states. } 
\label{hgb} 
\end{figure}

In the presence of  QCD instantons  ({\it a.k.a.}~monopoles)  quantization has to be performed in each topological sector.   In our phenomenological approach we approximate this by  generalizing the mode expansion of Eq.~(\ref{part}) to describe field fluctuations, $\bA_f$, with respect to  a classical background field, $\bA_B$.   
   \begin{equation} 
 \bA^a(\bx) \to \bA^a_f(\bx) + \bA_B^a(\bx). \label{shift} 
 \end{equation} 
This classical field mocks the nontrivial topological vacuum  and will be specified later.   Thus  Eq.~(\ref{part}) now applies to  $\bA_f \equiv   \bA -  \bA_B$ and  $\bm{\Pi_f} = \bm{\Pi}$. Since 
 Eq.~(\ref{shift}) is a canonical transformation the Hamiltonian can be obtained by substitution. 
 Thus in the background field, at large-R, the dominant contribution to the Coulomb interaction between quark and gluon charges is given by 
  
\begin{equation}  H_C  \to  H_{qq}  + H^D_{qg} + H^D_{gg} + H^D_{gb} + H^M_{gb}. \end{equation} 
 Here $H_{qq}$ is the interaction between quark charges mediated by the Coulomb potential, 
  \begin{equation} 
  H_{qq} = -\int d\bx d\by \rho_q(\bx,a) V_c(|\bx - \by|) \rho_q(\by,a) , 
\end{equation} 
$H^D_{qg}$ is the quark-gluon charge density interaction diagonal with respect to the gluon number, 
  \begin{equation} 
  H^D_{qg}   =   -\int d\bx d\by \rho_q(\bx,a) V_c(|\bx - \by|) \rho^D_g(\by,a) 
  \end{equation} 
  with  $\rho^D_g = \sum_i \rho^1_i(\bx,a) \alpha^{\dag}_i \alpha_i$, 
   and  $H^D_{gg}$ is the normal-ordered,  diagonal interaction between gluon charge densities 
\begin{equation} 
H^D_{gg} =  -:\int d\bx d\by \rho^D_g(\bx,a) V_c(|\bx - \by|) \rho^D_g(\by,a): .  \label{di}
\end{equation} 
Finally the two terms proportional to $\bA_B$,  $H^D_{gb}$ and $H^M_{gb}$, are given by 
 \begin{eqnarray} 
H^D_{gb}  +  H^M_{gb} = -\int d\bx d\by  \rho^B_g(\bx,a) V_c(|\bx-\by|) \rho^B_g(\by,a), \nonumber \\
\label{mix} 
\end{eqnarray} 
with 
\begin{equation} 
\rho^B_g(\bx,a) =  f_{abc} \bA_B^b(\bx) \bm{\Pi}^c(\bx) 
\label{sub} 
\end{equation} 
  and describe the interaction of physical gluons with the background field and 
 the  gluon pair creation in the presence of the background, respectively.
Physical states should be color neutral, thus creation or annihilation of a single gluon can be neglected. 
 In the presence of the background, the expectation value of the charge operator 
\begin{equation} 
Q^a_B =  f_{abc} \int d\bx  \bA_B^b(\bx) \bm{\Pi}^c(\bx) 
\end{equation} 
in physical states vanishes. However, in a simple classical model for the distribution of background fields, as described in Appendix~\ref{appa},  quantum charge fluctuations do not vanish, {\it i.e.}, $Q^aQ^a \ne 0$ even for color singlet states. We thus modify the right hand side of Eq.~(\ref{mix}) in such a way that these fluctuations do not contribute to the energy, yielding 
\begin{eqnarray} 
H^D_{gb}  +  H^M_{gb} & = &    - \int d\bx d\by \rho^B_g(\bx,a) V_c(|\bx-\by|)  \rho^B_g(\by,a) 
\nonumber \\ &   + &  V_c(0) Q^a_B Q^a_B.  
\label{mix-1} 
\end{eqnarray} 
After normal-ordering, the term in Eq.~(\ref{mix})  proportional to $\alpha^\dag\alpha$ defines $H^B_{gb}$, and the term proportional to $\alpha^\dag\alpha^\dag + h.c.$ gives $H^M_{gb}$.  The difference between the gluon density-density interaction and the normal ordered Hamiltonian of Eq.~(\ref{di})  is proportional to  either $\alpha^\dag \alpha$ or $\alpha \alpha + h.c$. These,  together  with the kinetic and magnetic terms combine 
 to~\cite{Szczepaniak:2001rg}  {\it  i) }renormalize $\omega$ via a gap equation which eliminates terms proportional to $\alpha \alpha + h.c$, and {\it  ii)} modify the single gluon energy.
 Thus the final Hamiltonian can be expressed in the form 
\begin{eqnarray} 
H  & \to &   H_g  + H_C \nonumber \\
 &= &  
 \sum_i E_i \alpha^\dag_i \alpha_i  + H_{qq}  + H^D_{qg} + H^D_{gg} + H^D_{gb} + H^M_{gb} 
 \label{heff} 
\end{eqnarray} 
where $E_i = E(k)$ is the single gluon energy in the presence of the background field. The action of these operators  on  gluon chain states is shown in Figs.~(\ref{Eq}), (\ref{hqq-self}), (\ref{hqq-qg}), and (\ref{hgb}).


\subsection{ The basis for the  gluon chain}

We  define the chain in a  large-$N_C$ limit by a model in which the gluon chain state  is a superposition of multi-gluon states, 
\begin{equation} 
|Q \bar Q, R \rangle   = \sum_N a_N |N \rangle 
\end{equation} 
with each state in the sum describing a product of $N$ single gluons ordered in color and 
space  along a straight light between the quark-antiquark sources,  
\begin{eqnarray}
 | N \rangle  & = &  \frac{Z_N}{{\cal V} \sqrt{N_C}} \int_{-R/2}^{R/2} dx_N\int_{-R/2}^{x_N} dx_{N-1}   \cdots  \int_{-R/2}^{x_{2}} 
dx_1 \nonumber \\
& \times &  Q^{\dag}_i\left(\frac{R}{2}\hat z \right) [G^\dag(x_N) \cdots G^{\dag}(x_1)]_{ij} \bar Q^{\dag}_j\left(-\frac{R}{2}\hat z\right)  |0\rangle.  \nonumber \\ \label{nc} 
\end{eqnarray}
In the large-$R$ limit the longitudinal, {\it i.e.}, along the $Q\bar Q$ axis, and perpendicular motions of gluons factorize. The spacial distribution of gluons  in the plane perpendicular to the $Q\bar Q$ axis is given by a single-particle wave function, $\psi(\bk,\lambda) =  e^{*i}(\bk,\lambda)\psi^i(\bk_\perp)$, which defines  the gluon operators, $G$, in the chain (${\bf x} = (0_\perp,x)$) 
\begin{equation} 
G^{\dag}_{ij}(x) =\sum_{\lambda} \int [d\bk] \alpha^{\dag}(\bk,\lambda,a) T^a_{ij}  \psi(\bk,\lambda)
e^{-i\bk \cdot \hat z x}. \label{chain} 
\end{equation}  
The normalization constant  $Z_N$ is obtained from $\langle N|N\rangle = Z^2_N (C_F I R)^N/\Gamma(N+1) = 1$  where $I$ is the normalization integral for the spacial wave function, $\psi$, 
($[d\bk_\perp] \equiv d^2\bk_\perp/(2\pi)^2$)
\begin{equation} 
I =  \langle \psi| \psi \rangle =  \int [d\bk_\perp] \psi^i(\bk_\perp) \delta_T^{ij}(\bk_\perp)  \psi^j(\bk_\perp).
\end{equation} 
In the large-$N_C$ limit, computation of the leading contributions to the matrix elements of the effective Hamiltonian of Eq.~(\ref{heff}) in the basis of the 
gluon chain states,  Eq.~(\ref{nc}),  is straightforward. 
The details and numerical results are presented in the next section. 

\section{ Formation of the Gluon Chain at large $Q{\bar Q}$ separation} 
\label{com}
As discussed in Sec.~\ref{vqq} one could consider two models for $V_c(R)$. In what we refer to as model-$I$ $V_c(r)$ will be linearly confining and of the form 
\begin{equation}
V^I_c(r) = \sigma_c r + V_c(0),  \label{lin} 
\end{equation}
and in model-$II$ $V_c$ is asymptotically flat, 
\begin{equation}
\lim_{r \to \infty} V^{II}_c(r) = V_c(\infty)    < \infty . \label{lin2} 
\end{equation}
 We concentrate on the 
 interactions induced by the effective Hamiltonian  in the limit of large quark-antiquark separation.

\subsection{Matrix elements of the effective Hamiltonian in the chain basis space} 
 
 The one body term, $H_g$, in Eq.~(\ref{heff}) acts independently on individual gluons  in the chain 
  created  by  the operators $G^\dag$ ({\it cf}. Eq.~(\ref{chain})). 
 Using 
 \begin{equation} 
 [G(x), G^{\dag}(y)]_{ij} =   C_F \langle \psi|\psi\rangle \delta(x - y)  = C_F I \delta(x-y),
 \end{equation} 
we find
\begin{eqnarray} 
 \langle N| H_g |N\rangle &=& Z^2_N (C_F \langle \psi|\psi\rangle)^{N-1} 
 \sum_{i=1}^N \int_{-R/2}^{R/2} dx_N \cdots  \int_{-R/2}^{x_{i+1}} dx_i  \nonumber \\
&  \times&  C_F \langle \psi |E |\psi \rangle   \int_{-R/2}^{x_i} dx_{i-1} \cdots \int_{-R/2}^{x_2} dx_1 \nonumber \\ 
 &  = & N \frac{  \langle \psi |E |\psi \rangle}{\langle \psi|\psi \rangle}\equiv N \frac{N_C}{2} [ e -    V_c(0)  ]  \label{1p} 
\end{eqnarray}
where 
\begin{equation} 
 \langle \psi |E |\psi\rangle = \int [d\bk_\perp] E(|\bk_\perp|) \psi^i(\bk_\perp) \delta_T^{ij}(\bk_\perp)  \psi^j(\bk_\perp).
 \end{equation}  
 and to define $e$  we subtracted from the single gluon energy a constant proportional to the negative of the potential at the origin. In color singlet states the total energy of the system should be invariant under a constant 
 shift~\cite{LeYaouanc:1987ff,LeYaouanc:1987bc}, which we now demonstrate. 
The  single gluon energy, $E(k)$,  contains self energies. In the variational approximation the component of the self energy due to the Coulomb interaction is given by~\cite{Szczepaniak:2001rg}
 \begin{equation} 
 \Sigma_C(k) = -\frac{N_c}{2} \int [d\bq] \tilde V_c(\bk - \bq)\frac{1 + \hat\bk\cdot \hat \bq}{2} \frac{\omega(k)}{\omega(q)} 
 \end{equation} 
 where $\tilde V_c$ is the Fourier transform of the Coulomb potential. For a linearly rising, confining potential, {\it e.g.}, model-$I$,  the low momentum singularity of $\tilde V(k)$ is not integrable and the resulting infinite self energy can be interpreted as a manifestation of confinement of color charges. A finite self energy is obtained by subtracting the IR singularity which leads to 
 \begin{equation} 
 \Sigma_C(k) = \Sigma'_C(k) - \frac{N_c}{2} V_c(0) 
 \end{equation}
 with $\Sigma'_C(k)$ finite and given by 
 \begin{equation} 
  \Sigma'_C(k) -\frac{N_c}{2} \int [d\bq] \tilde V_c(\bk - \bq)\left[ \frac{1 + \hat\bk\cdot \hat \bq}{2} \frac{\omega(k)}{\omega(q)} - 1\right] 
 \end{equation} 
that  follows from  
  \begin{equation} 
 V_c(0) = \int [d\bq] \tilde V_c(\bq). 
 \end{equation} 
 Even though  for a confining potential the Fourier transform is defined modulo a constant, it is expected that   when all, self and mutual, interactions between color charges are accounted for
   the dependence on $V_c(0)$ disappears from color singlet matrix  elements. This will also be the case for the matrix elements of the effective  Hamiltonian in the chain basis considered here. 
   In anticipation of this result, in Eq.~(\ref{1p}) we defined an IR finite single particle energy $e(k)$ by subtracting the Coulomb self energy equal to $-N_c V_c(0)/2$. 
 Thus, in the last line of Eq.~(\ref{1p}), $e$ is finite, and the IR singularity of the  confining Coulomb potential is explicit in the term proportional to $V_c(0)$.  In the case of model-$II$ with non-confining interactions, self-energies are IR finite but we can perform the subtractions nevertheless.

\begin{figure}[t!]
\centerline{\scalebox{0.30}{\includegraphics{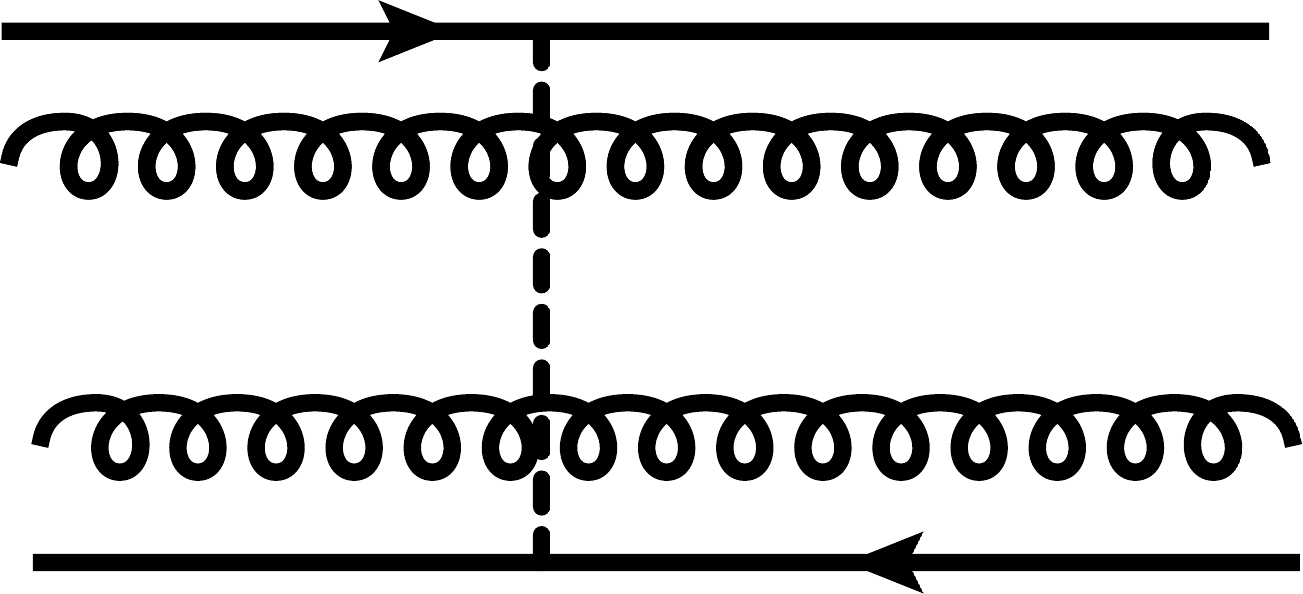}}}
\caption{ A non-planar diagram, which we do not take into account, representing direct Coulomb interaction
 between the quark and the antiquark in the presence of $2$ gluons. } 
\label{hqq-exchange}
\end{figure}

 In the absence of  chained gluons, $N=0$, the interaction between quark densities produces the Coulomb potential between quark charges ({\it cf}. Eq.~(\ref{neq0})). With $N$ gluons separating 
the quark from the  antiquark,  the direct interaction between quark charges is nonplanar ({\it cf}.  Fig.~(\ref{hqq-exchange})) and suppressed by a power of $N_C$ compared to successive Coulomb interactions between the quark and the nearest gluon or  the interaction between any two nearest-neighbor gluons in the chain. To leading order in $N_C$ the $H_{qq}$  contribution 
 thus reduces to the quark self energies,

\begin{equation} 
 \langle N| H_{qq} |N\rangle  = - C_F V_c(0) \to -\frac{N_C}{2} V_c(0).  \label{e1} 
 \end{equation} 
 The quark gluon interaction, to leading order in $N_C$, couples the quark (or the antiquark) to the nearest gluon in the chain. For example, for the antiquark-gluon interaction we find 
 
 \begin{eqnarray} 
 \langle N|H^D_{qg} |N\rangle &\to& Z_N^2  (C_F \langle \psi |\psi\rangle )^{N-1}  \nonumber \\
& \times & \int_{-R/2}^{R/2} dx_N \cdots \int_{-R/2}^{x_2} dx_1 C_F \langle g\bar Q| H_{qg}(x_1) |g\bar Q\rangle  \nonumber \\
& = &  \frac{N}{R^N} \int_{-R/2}^{R/2} dx_1 \left(x_1 + \frac{R}{2}\right)^{N-1} 
\frac{ \langle g\bar Q| H_{qg}(x_1) |g\bar Q\rangle }{\langle \psi|\psi \rangle}  \nonumber \\
\end{eqnarray}
where 
\begin{eqnarray} 
& &  \langle g\bar Q| H_{qg}(x_1) |g\bar Q\rangle =  \frac{N_C}{2} \int d^2\bx_\perp   [d\bk_\perp] [d\bq_\perp] 
  e^{i(\bk_\perp-\bq_\perp)\cdot\bx_\perp} \nonumber \\
  & & \times \frac{1}{2} \left[ \sqrt{\frac{\omega(|\bk_\perp|)}{\omega(|\bq_\perp|)} } 
  +  \sqrt{\frac{\omega(|\bq_\perp|)}{\omega(|\bk_\perp|)} }  \right]
V_c\left(\sqrt{ |\bx_\perp|^2 + |\frac{R}{2} - x_1|^2} \right)  \nonumber \\   
& &  \times   \psi^i(\bk_\perp) [\delta_T(\bk_\perp)\delta_T(\bq_\perp)]^{ij}\psi^j(\bq_\perp). \nonumber \\
 \end{eqnarray}
In the limit $R \to \infty$ where $x_1/R = O(1)$  this reduces to 
\begin{equation} 
 \langle g\bar Q| H_{qg}(x_1) |g\bar Q\rangle  = \frac{N_C}{2}  V_c\left(|\frac{R}{2} - x_1|\right)  \langle \psi|\psi \rangle.
 \end{equation} 
Taking into account both quark and antiquark contributions, for the $H_{qg}^D$ matrix element
we  obtain 
\begin{equation} 
 \langle N|H^D_{qg} |N\rangle = N  N_C \int_{-1/2}^{1/2} dz \left(z + \frac{1}{2}\right)^{N-1} 
 V_c\left(R\left(z - \frac{1}{2}\right)\right). \label{hdqg} 
\end{equation} 
For the linearly rising potential  of model-$I$ 
Eq.~(\ref{hdqg}) yields 
\begin{equation} 
 \langle N|H^D_{qg} |N\rangle^I =  N_C \frac{\sigma_c  R}{N+1}  + N_C V_c(0). \label{e2-I}
 \end{equation} 
 while in the case of model-$II$ we find 
 \begin{equation} 
 \langle N|H^D_{qg} |N\rangle^{II}  =  N_C V_c(\infty). \label{e2-II}
 \end{equation} 
 The interaction between two nearby gluons in the chain given by $H^D_{gg}$ is also straightforward to compute, and passing directly to the $R \to \infty$ limit we find,  
 \begin{eqnarray} 
&& \langle N|H^D_{gg} |N\rangle  =  Z_N^2 (C_F \langle \psi |\psi\rangle )^{N-2} \sum_{i}^{N-1} \int_{-R/2}^{R/2} dx_N \cdots \nonumber \\
 &  \times &  \int^{x_{i+2}}_{-R/2} dx_{i+1} \int_{-R/2}^{x_{i+1}}dx_i  C_F \left(\frac{N_C}{2}\right)^2 V_c(x_{i+1} - x_i) \langle\psi|\psi\rangle^2\nonumber \\
 & &\times   \int_{-R/2}^{x_i} dx_{i-1} \cdots \int_{-R/2}^{x_2} dx_1 \nonumber\\ 
& &   = \frac{N!}{C_F}  \int_{-1/2}^{1/2} dz \int_{-1/2}^z dw \left( \frac{N_C}{2}\right)^2 
 V_c(R(z - w)) \frac{(1 + w -z)^{N-2}}{(N-2)!}.  \nonumber \\
\end{eqnarray}
For the linear potential of Eq.~(\ref{lin}), to leading order in $N_C$ this yields 
\begin{equation}
 \langle N|H^D_{gg} |N\rangle^I  = \frac{N_C}{2} \frac{N-1}{N+1} \sigma_C R +   \frac{N_C}{2} (N-1) V_c(0) \label{e3-I}
\end{equation}
and for the asymptotically constant potential 
\begin{equation}
 \langle N|H^D_{gg} |N\rangle^{II}  = \frac{N_C}{2} (N-1 ) V_c(\infty). \label{e3-II}
\end{equation}
Since all terms in the effective Hamiltonian (including the self energies)  are $O(g^2)$, 
 and   $\lim_{N_C \to \infty} N_C g^2 = O(1)$,  at large $N_C$  all matrix elements are 
  finite when expressed in terms of   $\bar e = N_Ce/3 $,  $\bar \sigma_c = N_C \sigma_c/3$ for model-$I$ 
   and $\bar V_c(\infty) = N_C V_c(\infty)/3$, $\bar V_c(0) \equiv N_C V_c(0)/3$ for model-$II$, respectively.

   Adding all  diagonal contributions of the effective Hamiltonian matrix that   
 are independent of the background field, 
 we thus find,
\begin{equation} 
\langle N|H|N \rangle^I = \frac{3}{2} N \bar e  + \frac{3}{2} \bar \sigma_C R   \label{diag-I} 
\end{equation} 
and 
\begin{eqnarray} 
\langle N|H|N \rangle^{II}  &= &  \frac{3}{2} N \bar e  + \frac{3}{2}  (N + 1) ( \bar V_c(\infty)  - \bar V_c(0))  \nonumber \\
& \equiv  & \frac{3}{2} N m_g + c     \label{diag-II}  
\end{eqnarray} 
for model-$I$ and model-$II$,  respectively. 
 For $N=0$ this agrees with Eq.~(\ref{neq0}), while, for $N\ge 1$, eigenstates of 
   Eq.~(\ref{diag-I})  or (\ref{diag-II}) represent a tower of chain states with energies proportional to the number of gluons in the chain. Clearly the lowest energy state of the diagonal part of the Hamiltonian 
    is the variational $Q\bar Q$ state, with $N=0$ gluons. 
 The genuine chain contribution to the lowest energy state must therefore originate from 
  the terms in the Hamiltonian which couple the constituent gluons with the background field, as expected. 
The interaction of physical gluons with the background is given by 


\begin{eqnarray} 
& & \langle N| H^D_{gb} |N\rangle  =   Z_N^2 (C_F \langle \psi|\psi\rangle)^{N} \sum_{i=1}^{N-1} \int_{-R/2}^{R/2} dx_N \cdots  \int_{-R/2}^{x_{i+2}} dx_{i+1} \nonumber \\
& \times & \int_{-R/2}^{x_{i+1}} dy_i \int_{{-R/2}}^{y_{i}} dx_i  F_B(|y_i - x_i|) \int_{-R/2}^{x_i} dx_{i-1}   \cdots \int_{-R/2}^{x_2} dx_1 \nonumber \\
\end{eqnarray}

 where 
 
 \begin{equation} 
 F_B(|y-x|) = \frac{N_C}{2} \gamma(|y-z|) (V_c(|y-z|)-V_c(0)),  \label{gamma} 
 \end{equation}
 and  
 \begin{eqnarray} 
& & \gamma  =    \int d\bx_\perp d\by_\perp [d\bk_\perp][d\bq_\perp]  \sqrt{\omega(\bk_\perp) \omega(\bq_\perp) } e^{-i \bq_\perp \bx_\perp + i \bk_\perp \by_\perp}  \nonumber \\
& & \times  \frac{[ \psi(\bq_\perp) \delta_T(\bq)]^i G^{ij}_c(\bx_\perp-\by_\perp,y-z)  [ \psi(\bk_\perp) \delta_T(\bk)]^j }{\langle \psi| \psi\rangle}. \label{hdd} 
\end{eqnarray}
The correlation function  $G_c$ is obtained from the density of the vacuum fields 
  \begin{equation} 
G^{ij}(\bx_\perp - \by_\perp,x-y)  = \frac{\langle A^{ia}_c(\bx_\perp,x)  A^{ja}_c(\by_\perp,y) \rangle}{N^2_C-1}.  \label{g} 
\end{equation}  
  Here the expectation value is taken with respect to the distribution of sources of the background field.   These might effectively describe monopole-antimonopole pairs in 3D,  vortex surfaces in 4D, merons, {\it etc.}.  A simple model is considered in the Appendix. 
     Since it is these background fields that are responsible for confinement in the first place, {\it i.e.} generation of the Coulomb potential $V_c(R)$, we assume that the  density of  the underlying magnetic sources is approximately uniform over the quark-antiquark separation. So for $  |x-y| \lsim R$ we expect in general   
     \begin{equation} 
G^{ij}(\bx_\perp - \by_\perp,x-y)  \sim  G^{ij}(\bx_\perp - \by_\perp), 
\end{equation}  
  and $\gamma$ in Eq.~(\ref{gamma}) reduces to a constant 
    of $O(\Lambda_{QCD})$, {\it i.e.} it is independent of the longitudinal distribution of gluons along the chain. 

For model-$I$   evaluation of the integrals in Eq.~(\ref{hdd})  then gives 
\begin{equation}
\langle N |H^D_{gb}|N\rangle^I = N_C N! \frac{N-1 }{\Gamma(N+3)} \gamma \sigma_C   R^2
 = 3 \frac{N-1}{(N+1)(N+2)} \gamma \bar \sigma_C R^2.  \label{b1-I}
 \end{equation} 
while for model-$II$ we find 

\begin{equation}
\langle N |H^D_{gb}|N\rangle^{II}  
 = 3 \gamma R   \frac{N-1}{N+1}    (\bar V_c(\infty) - \bar V_c(0) ).  \label{b1-II}
 \end{equation} 

Finally we consider the components of the interaction between physical gluons and the background that changes the gluon number.  
From Eq.~(\ref{mix})  we find (for $N\ge 2$) 

\begin{eqnarray}
 & &  \langle N-2 | H^M_{gb}|N\rangle = \langle N|H^M_{gg}|N-2\rangle = 
  Z_{N} Z_{N-2} (C_F \langle \psi|\psi\rangle)^{N-1} \nonumber  \\
  & & \times \sum_{i=1}^{N-1} \int_{-R/2}^{R/2} dx_N \cdots 
\int_{-R/2}^{x_{i+1}} dx_i F_B(x_{i+1}-x_i) \cdots \int_{-R/2}^{x_2} dx_1 \nonumber \\
  \end{eqnarray} 
  
  which gives 
  \begin{equation}
  \langle N-2 | H^M_{gb}|N\rangle^I = \langle N|H^M_{gg}|N-2\rangle^I = 
  \frac{3 \sqrt{N(N-1)}}{2N(N+1)} \gamma \bar \sigma_c R^2. \label{b2-I}
  \end{equation}
and 
  \begin{eqnarray}
  \langle N-2 | H^M_{gb}|N\rangle^{II} & = &  \langle N|H^M_{gg}|N-2\rangle^{II}  =  \nonumber \\
 & = &   \frac{3}{2} \gamma R \sqrt{\frac{N-1}{N}}
   (\bar V_c(\infty) - \bar V_c(0) ). \label{b2-II}
  \end{eqnarray}
for the two models, respectively. 

Collecting all the terms, Eqs.~(\ref{diag-I}),(\ref{b1-I}),(\ref{b2-I}) for model-$I$ and Eqs.~(\ref{diag-II}),(\ref{b1-II}),(\ref{b2-II}) for model-$II$,   we find the following expression for the matrix elements of the Hamiltonian in the gluon chain basis for large-$N$,


\begin{eqnarray} 
\langle N' | H | N\rangle^I & =  & \frac{3}{2} N \bar e\delta_{N'N} + \frac{3}{2}  R\bar \sigma_C  \left(1 
 + r \frac{R}{N} \gamma  \right) \delta_{N'N} \nonumber \\
  & + & \frac{3}{2}   \gamma \bar \sigma_C  \frac{R}{N}  R   \delta_{N',N-2} 
  +  \frac{3}{2}  \gamma \bar \sigma_C \frac{R}{N'} R   \delta_{N'-2,N} \label{fin-I} 
  \end{eqnarray}

\begin{eqnarray} 
\langle N' | H | N\rangle^{II} & =  & \frac{3}{2} N \bar m_g \delta_{N'N} +  \frac{3}{2} r  \gamma  R 
(\bar V_c(\infty) - \bar V_c(0))   \delta_{N'N} \nonumber \\
  & + & \frac{3}{2}    \gamma   R 
  (\bar V_c(\infty) - \bar V_c(0))  (   \delta_{N',N-2} 
 +  \delta_{N'-2,N} )  \nonumber \\ \label{fin-II} 
  \end{eqnarray}
Here $r$ is the ratio of the diagonal to off-diagonal matrix elements in the limit of large-$R$. The specific value $r=2$ follows from the fact that in the two models both terms originate from the same interaction {\it cf}. Eq.~(\ref{mix-1}). 
Below, while presenting numerical result, we will also discuss the dependence of the lowest eigenvalues on this ratio.

 \subsection{Numerical Results} 
 
\begin{figure}[t!]
\centerline{\scalebox{0.450}{\includegraphics[angle=-90]{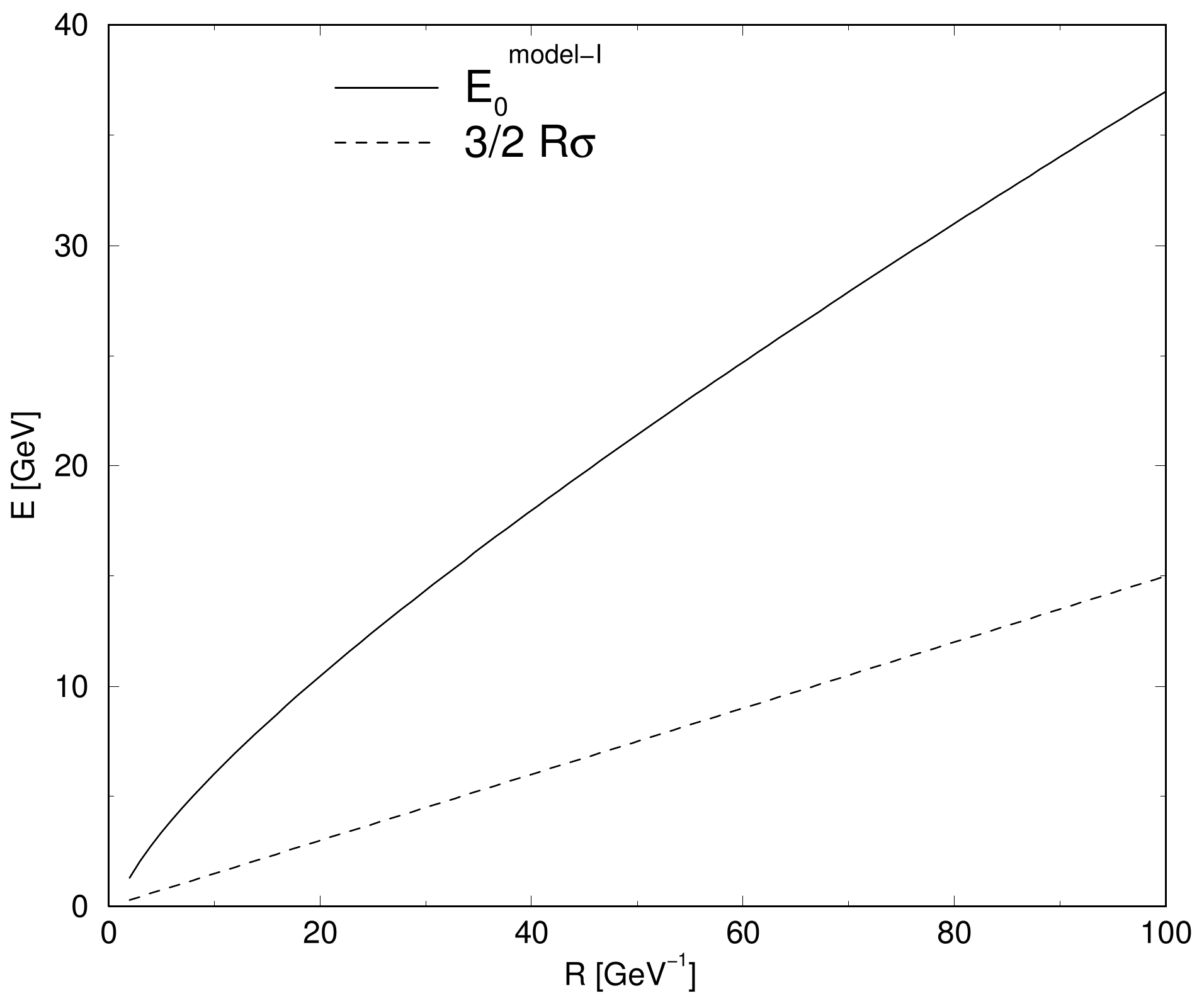}}}
\caption{ Ground state energy (solid line) of the chain Hamiltonian in model-$I$. A power  law fit yields $E^I_0 = 
0.984 (R GeV)^{0.787} GeV$. The dashed line gives the energy of the bare state using for the string tension 
   $\bar \sigma_C = 0.1\mbox{ GeV}^2$. }
   \label{E-I}
\end{figure}

\begin{figure}[t!]
\centerline{\scalebox{0.450}{\includegraphics[angle=-90]{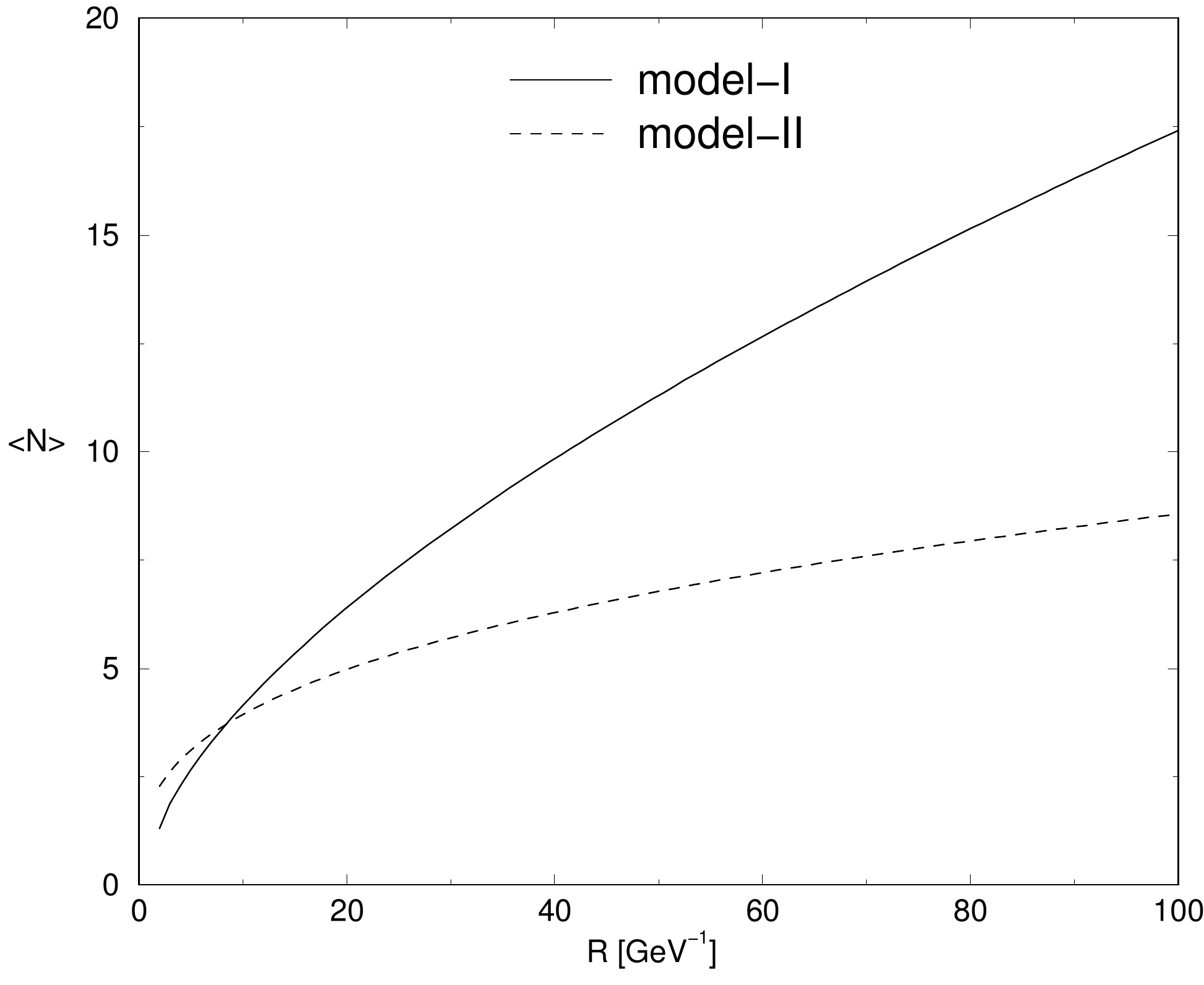}}}
\caption{ Expectation value of the the number of gluons in the chain as a function of $Q{\bar Q}$ separation, $R$. 
 A power law fit gives $\langle N \rangle^I = 0.985 (R \mbox{ GeV})^{0.623}$ and 
 $\langle N \rangle^{II}  = 0.984 (R \mbox{ GeV})^{0.787}$ for model-I (solid line) and model-II (dashed line), respectively. }
   \label{N}
\end{figure}

Before analyzing the spectra of the effective chain  model Hamiltonians we consider the large-$R$ limit of  the matrix 
\begin{equation} 
\langle N' | H| N \rangle = R ( \delta_{N',N-2} + \delta_{N'-2,N} ). 
\end{equation} 
It is straightforward to show that the ground state energy of $H$ for large-$R$ is $-2R$.  For the Hamiltonian of 
 model-$I$ this implies that if the kinetic term  (proportional to $N \bar e$) was ignored, the lowest eigenvalue 
  of $H^I$ for large-$R$ would behave as 

\begin{equation} 
\frac{3}{2}\gamma \bar \sigma_C \left(1 + r\frac{R}{\langle N\rangle} \right) - 2 \frac{3}{2} \gamma \bar \sigma_C 
  \frac{R}{\langle N\rangle}  = \frac{3}{2} \gamma  \bar \sigma_C R + \frac{3}{2} \gamma \bar \sigma_C (r-2) \frac{R^2}{\langle N\rangle}. 
  \end{equation}
 At $r=2$ the quadratic term vanishes and the lowest chain state energy is expiated to grow linearly with $R$. 
  At large-$R$ if
    $r >> 2$  then  the lowest eigenvalue is dominated by the diagonal term. In this case the expectation value of $N$, 
\begin{equation} 
\langle N \rangle = \frac{\sum_{N} N  |\psi_0(N) |^2 }{\sum_N |\psi_0(N)|^2}, 
\end{equation} 
where $\psi_0$ is the wave function of the lowest energy chain state,  can be determined by minimizing the diagonal part with respect to $N$. This gives 
\begin{equation} 
\langle N \rangle = \sqrt{\frac{2 \gamma}{\bar e}} R 
\end{equation} 
and the ground state energy approaches 
\begin{equation} 
E_0^{I} = 3 R \sqrt{r \gamma \bar \sigma_C \bar e}  + \frac{3}{2} R\bar \sigma_C .
\end{equation} 
Thus for $r>2$ the energy of the chain is higher than the energy of the bare state,  $|0\rangle$. 
If $r< 2$  the off-diagonal term dominates  and the ground state energy becomes negative and proportional to $ -R^2$ while the average number of gluons in the chain  $\langle N \rangle = O(1)$. 
However, when the kinetic term is included  in the critical case $r=2$ the lowest energy of the chain state no longer increases linearly with $R$. After numerical diagonalization 
we find 
\begin{equation} 
 \langle N \rangle^I \propto (R \mbox{ GeV})^{0.623 \pm 0.004}, \\ 
\frac{ E_ 0^I }{\mbox{GeV}} \propto (R \mbox{ GeV})^{0.787 \pm 0.006} 
\end{equation} 
for  a typical  set of parameters $\bar e = 600 \mbox{ MeV}$, $\gamma = 1\mbox{ GeV} $ and $\bar \sigma_C = 0.1\mbox{ GeV}^2$, and we find weak dependence of the exponents  on these parameters. 
That is, for the chain model-$I$, we find that the lowest energy chain state has higher energy than the bare state. 
 In the critical case the energy increases  less rapidly than the length of the chain, $R$, and  is proportional to $R^2$    for $r>2$.  The average number of gluons grows weakly with $R$. The results are summarized in Figs.~(\ref{E-I}),~(\ref{N}) .

\begin{figure}[t!]
\centerline{\scalebox{0.450}{\includegraphics[angle=-90]{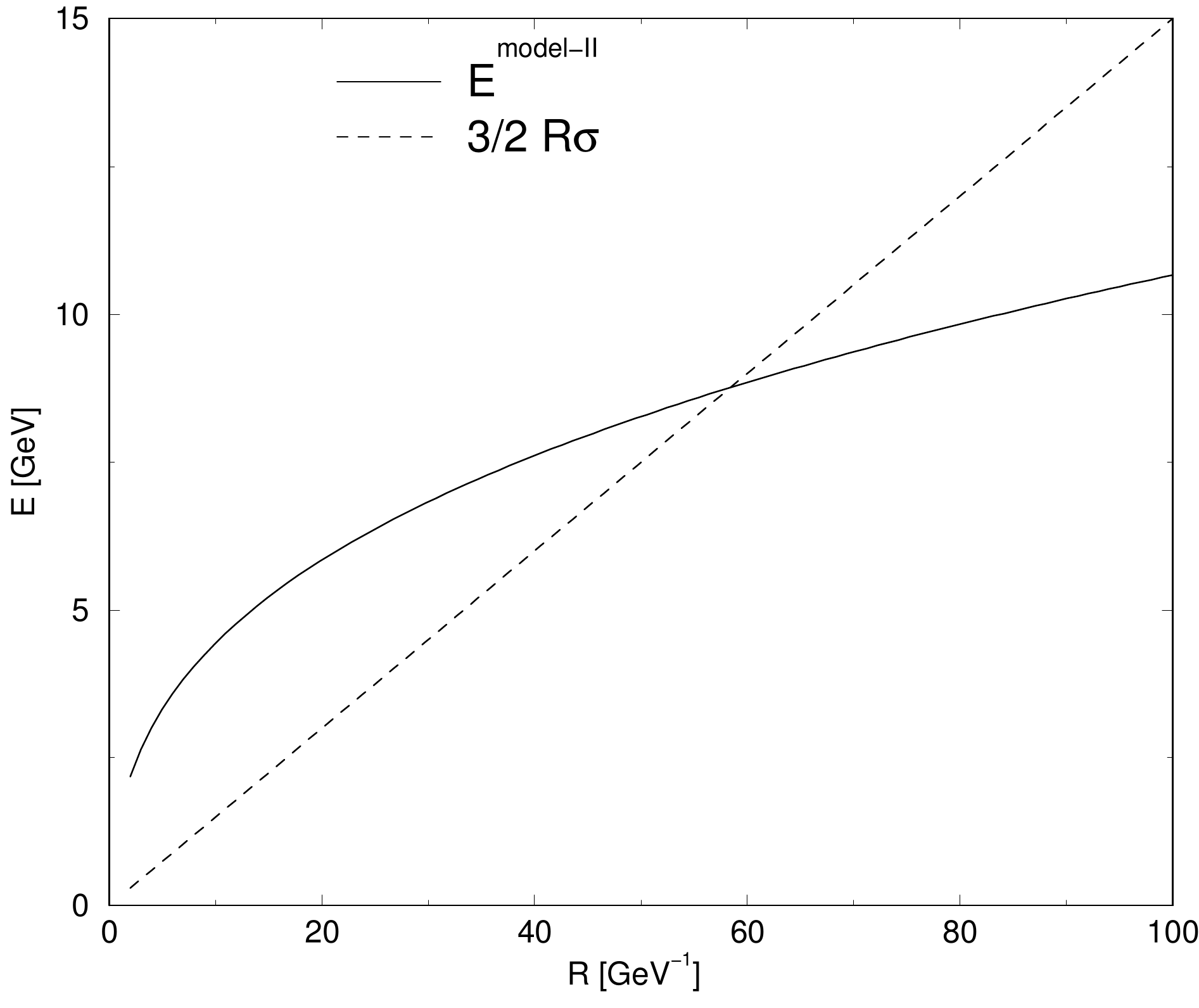}}}
\caption{ Ground state energy (solid line) of the chain Hamiltonian in model-$II$. A power  law fit yields $E^I_0 = 
1.867 (R GeV)^{0.379} GeV$. The dashed line gives the energy of the bare state using for the string tension 
   $\bar \sigma_C = 0.1\mbox{ GeV}^2$. }
   \label{E-II}
\end{figure}

In the case of model-$II$ for $r > 2$,  one easily finds,  
 \begin{eqnarray}
 E_0^{II}  &=& \frac{3}{2}(r-1) \gamma R[\tilde V_c(\infty) - \tilde V_c(0)] \nonumber \\
 \langle N \rangle^{II} & \propto & R^{1/3} 
 \end{eqnarray}
 while for $r<2$ with the off-diagonal term dominating, 
 \begin{equation} 
E_0^{II} = -\frac{3}{2}(r-1) \gamma R[\tilde V_c(\infty) - \tilde V_c(0)]. 
\end{equation} 
Finally for the critical choice $r=2$  numerical digitalization  yields 

\begin{equation} 
 \langle N \rangle^{II} \propto (R \mbox{ GeV})^{0.338 \pm 0.005}, \\ 
\frac{ E_ 0^{II} }{\mbox{GeV}} \propto (R \mbox{ GeV})^{0.379 \pm 0.005} 
\end{equation} 
for the set of parameters, $m_g = 600 \mbox{ MeV}$, $\bar \sigma_C = 0.1\mbox{ GeV}^2$, and $\gamma = 1\mbox{ GeV}$, $\hat V_c(\infty) - \hat V_c(0) = 1 \mbox{ GeV}$, and the results are 
 shown in Fig.~(\ref{N}),~(\ref{E-II}).  
 
In model-$II$  as $R$-increases at some point the energy of the ground state chain increases less than the Coulomb potential. The chain state, however, has energy which is higher than that of the bare state, with the latter approaching a constant at large-$R$. Thus in both models interactions among the chain increase the energy of the $Q{\bar Q}$ pair as compared to the state with no gluons.


 \section{Summary and Outlook} 
 \label{last} 
We investigated microscopic origins of the gluon chain model. By analyzing the physical gauge interactions among 
 constituent gluons, we fund a scenario for generating a chain. In this scenario a state with a number of gluons in the chain that is  increasing with the separation between the $Q{\bar Q}$ source emerges from interactions of dynamical gluons with the background field. The background field is necessary in a phenomenological model of confinement if the latter is to originate from condensation of chromomagnetic charges. 
These interactions introduce off-diagonal elements into the effective Hamiltonian,  which is one of the main differences between this and the chain model where the pair-production is absent.  We have shown that the resulting ground state energy is convex~\cite{bachas}  but the two models considered are still too simplistic to generate the linearly rising potential. While this deficiency can potentially be improved by considering more sophisticated models for the background field we found it difficult to reproduce the Zwanziger conjecture of "no-confinement without Coulomb confinement" ~\cite{Zwanziger:2002sh}. We find the energy of the chain state to be higher then that of the bare one, defined as the expectation value of the Coulomb kernel in a state with no-backward reaction from the sources on the vacuum. It is possible that  a resolution of this problem  requires  renormalization for the single-gluon energies in the presence of the chain so that effectively $\bar e$ decreases with the number of gluons.



\acknowledgments{This research is supported in part by INFN and  
the U.S.\ Department of Energy under Grant No.\ 
DE-FG0287ER40365. A.O also acknowledges support from the NSF-sponsored Summer Research Experience for Undergraduate  (RUE)  program at Indiana University PHY-1156540.  } 

\appendix

\section{Background field model} 
\label{appa}
The correlation function $G$ defined in Eq.~(\ref{g})  is computed using a classical distribution of sources of the background field. For example if these are monopole- antimonopole pairs the density 
$\rho_N(\bc_i \bar \bc_i)$ depends on the locations  of the $N$ pairs. 
 The expectation value of a function of $\bA_c$ is computed 
 from 
\begin{equation} 
\langle \bA^a_c(\bx) \bA^a_c(\by) \rangle =\frac{Z[\bA_c \bA_c]}{Z[1]} 
\end{equation} 
where 
\begin{equation} 
Z[{\cal O}[\bA_c]] = \sum_{N}^\infty 
\int dn^a \int \Pi_i^N d\bc_i  d\bar \bc_i
 \rho(\bc_i \bar \bc_i)
  {\cal O}[\bA_c]
 \end{equation} 
and the background field is given by 
 \begin{equation} 
 \bA_c^a(\bx) = n^a  \sum_{i=1}^{N}  [ \bA_m(\bx_\perp - \bc_{\perp,i}) -    \bA_m(\bx_\perp - \bar\bc_{\perp,i})] \label{m} 
 \end{equation} 
where $\bA_m$ is the abelian monopole field, and $n^a$ represents the (common) orientation of monopoles in the $SU(N_C)$ algebra.  For a uniform distribution of monopole-antimonopole  pairs 
 along the $Q\bar Q$-axis  ($\hat z$-axis) with the density given by 
    \begin{equation} 
\rho(\bc_i,\bar \bc_i) = \frac{\rho^N}{ ({\cal V}_\perp R)^NN!} \Pi_{i=1}^N \theta(\frac{R}{2} -  |\bc_i \hat z|)  
 \theta(\frac{R}{2} -  |\bar \bc_i \hat z|)  \label{rr}
\end{equation} 
 the background field is approximately constant along the $Q\bar Q$ axis. 
In Eq.~(\ref{rr}) $\rho$ is the density of  monopoles which is equal to the density of antimonopoles
\begin{equation} 
\rho = \frac{\rho_\perp (N_C^2-1)}{R} . \label{ds} 
\end{equation} 

 For the correlation function $G(\bx_\perp,x) $ we then obtain 

\begin{eqnarray} 
G(\bx_\perp,x) = G(\bx_\perp)   & = &   \rho_\perp  \left[ \int d\bc_\perp  \bA_m(\bx_\perp- \bc_\perp) \bA_m(\bc_\perp)  
 \right. \nonumber \\
 &  -  &   \left.  \frac{1}{ {\cal V}_\perp} \int d\bc_\perp d\bc'_\perp  \bA_m(\bc_\perp) \bA_m(\bc'_\perp)  \right]. \nonumber \\
\end{eqnarray} 
The last term  originates from the charge neutrality of the monopole-antimonopole distribution. 
If the core of the monopole  field is smoothed out over a distance scale $a = O(\Lambda^{-1}_{QCD})$ then 
\begin{equation} 
G(\bx_\perp) \sim  \rho_\perp \log  \frac{| \frac{R}{2} - |\bx_\perp| |}{a} 
\end{equation} 
where the $\log R$ dependence comes from cutting off the long range integral over the transverse plane. This is the standard expression for the correlation function of a pair of  2D vortices  separated by a distance $R$. The 2D reduction originates from the assumption the monopoles are uniformly distributed, Eq.~(\ref{ds}),   along the direction of the  $Q\bar Q$ separation.

\end{document}